\documentclass[11pt,twoside]{article}
\usepackage{asp2010}

\resetcounters

\newcommand{\msun}{M$_\odot$}

\begin{document}

\title{The origin of very wide binary stars}
\author{M.B.N. Kouwenhoven$^1$, S.P. Goodwin$^2$,  Melvyn B. Davies$^3$,
  Richard~J.~Parker$^{4,2}$, P.~Kroupa$^5$, and D. Malmberg$^3$
\affil{$^1$Kavli Institute for Astronomy and Astrophysics, Peking
  University, Yi~He~Yuan~Lu~5, Haidian District, Beijing 100871, P.R. China}
\affil{$^2$Department of Physics and Astronomy, University of Sheffield, Hicks Building, Hounsfield Road,
  Sheffield S3 7RH, United Kingdom}
\affil{$^3$Lund Observatory, Box 43, SE-221 00, Lund, Sweden}
\affil{$^4$Institute for Astronomy, ETH Z\"{u}rich, Wolfgang-Pauli-Strasse 27, 8093 Z\"{u}rich, Switzerland}
\affil{$^5$Argelander Institute for Astronomy, University of Bonn, Auf
dem H\"{u}gel 71, 53121 Bonn, Germany}}

\begin{abstract}
  A large population of fragile, wide ($>10^3$~AU) binary systems
  exists in the Galactic field and halo. These wide binary stars
  cannot be primordial because of the high stellar density in star
  forming regions, while formation by capture in the Galactic field is
  highly improbable. We propose that these binary systems were formed
  during the dissolution phase of star clusters \citep[see][for
  details]{kouwenhoven2010}. Stars escaping from a dissolving star
  cluster can have very similar velocities, which can lead to the
  formation of a wide binary systems. We carry out $N$-body
  simulations to test this hypothesis. The results indicate that this
  mechanism explains the origin of wide binary systems in the
  Galaxy. The resulting wide binary fraction and semi-major axis
  distribution depend on the initial conditions of the dissolving star
  cluster, while the distributions in eccentricity and mass ratio are
  universal.  Finally, since most stars are formed in (relatively
  tight) primordial binaries, we predict that a large fraction of the
  wide ``binary stars'' are in fact higher-order multiple systems.
\end{abstract}

\section{Wide binary systems in the Galactic field and halo}

The large majority of stars are thought to form as part of a binary or
multiple stellar system \citep[e.g.,][]{duquennoy1991, fischer1992,
  goodwinkroupa2005,kouwenhoven2005,kouwenhoven2007}. The general
consensus is that most star form in embedded star clusters and
loosely-bound associations \citep[e.g.,][]{lada2003,bastian2011},
which initially exhibit a significant amount
of substructure \citep[e.g.,][]{allison2009}. Following proto-star formation, the
properties of the binary population evolve over time, primarily due to
the effects pre-main sequence evolution \citep{kroupa1995} and
dynamical interactions with other stars
\citep[e.g.,][]{heggiehut, marks2011}. Most star clusters
dissolve within $10-50$~Myr after their formation \citep[see][and
references therein]{grijs2007}. The field star population is therefore
thought to be
the result of a mixture of stars originating from different star
clusters \citep{goodwin2010}.

\begin{figure}[tb]
  \begin{center}
  \includegraphics[width=\textwidth, height=!]{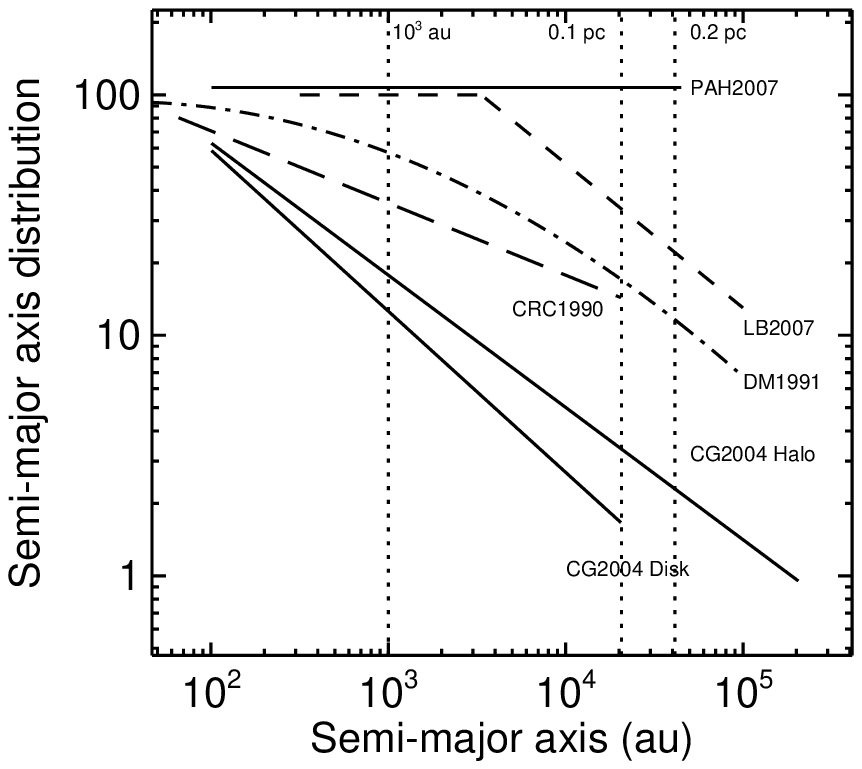}
  \caption{The observed semi-major axis distribution for wide binary
    systems, compiled from the catalogues of \cite{duquennoy1991,close1990,
      lepine2007,chaname2004} and \cite{poveda2007}.  \label{fig:kouwenhoven1}}
  \end{center}
\end{figure}

Over the last decades a significant number of wide ($>10^3$~AU) binary
systems have been discovered (see Fig.~\ref{fig:kouwenhoven1}). In the
log-normal period distribution of \cite{duquennoy1991}, for example,
$\sim 15\%$ of the binary systems have a semi-major axis larger than
$10^3$~AU. Individual wide binary systems are often identified in
proper motion studies, and occasionally combined with parallaxes,
radial velocity measurements and background star statistics
\citep[e.g.,][and numerous others]{makarov2008, quinnsmith,
  shaya2011}. The overall properties of the wide binary population can
also be obtained statistically \citep[e.g.,][]{longhitano2010}.  Wide
binary systems are extremely fragile, and those wider than
$0.1-0.2$~pc are easily destroyed in the Galactic field (see
Fig.~\ref{fig:kouwenhoven1}). This upper limit can be explained by
interactions with other stars, molecular clouds, and the Galactic
tidal field \citep[e.g.,][]{retterer1982, jiang2010}.  The properties
of wide systems in the Galactic field are also used to constrain the
properties of hypothesized dark components
\citep[e.g.,][]{quinn2009,allen2007,hernandez2008}.

Wide binary systems\footnote{Following \cite{kouwenhoven2010}, we
  define systems with an orbital separation in the range $10^3$~AU $\leq a \leq
  0.1$~pc as wide binary systems (or wide multiple systems).} cannot
have formed as primordial binaries in star clusters, simply because
their orbital separation is comparable to the size of a typical
embedded cluster. Moreover, the typical size of a star forming core
is $\sim 10^4$~AU \citep{ward-thompson2007}, which sets an absolute maximum to
the size of a primordial wide binary system. Even if they were somehow able to form, they would
be destroyed immediately due to stellar encounters
\citep{kroupa2001,parker2009}. 

The fact that it is not possible to form primordial binary systems with
semi-major axes in the range $10^3$~AU$-0.1$~pc implies that
wide binaries are formed at a later stage, as a result of dynamical
interactions between stars. Energy conservation implies that two stars
on an initially unbound orbit will remain unbound. Capture is
therefore only possible when energy is removed from the system, for
example by a third star that is present during the encounter. The formation rate $\dot{N}_B$ of binary systems via
three-body encounters is given by
\begin{equation}
  \dot{N}_B = 0.75\,\frac{G^5M^5n^3}{\sigma^9} \ ,
\end{equation}
\citep{goodman1993}, where $G$ is the gravitational constant, $M$ is
the typical stellar mass, $n$ the stellar number density, and $\sigma$
the velocity dispersion. The value of $\dot{N}_B$ is negligible for
stars in the Galactic field and halo. On the other hand, capture is
possible in the dense cores of star clusters, but this will never
result in the formation of long-lived {\em wide} binary systems, due
to the crowded stellar environment. 

In \cite{kouwenhoven2010} we proposed that wide binary systems form
during the dissolution phase of star clusters. This mechanism can
result in a significant population of binary systems \citep[see
also][]{moeckel2010, moeckel2011}.  In this scenario, an unbound pair
of escaping stars can form a binary system when their relative
velocity is small\footnote{A large population of comets may also be
  captured by stars via a similar mechanism, during cluster dissolution
  \citep[e.g.,][]{eggers1997,levison2010}}.  Our $N$-body simulations
(see below) result in a population of wide binary systems with
semi-major axes comparable to their initial separation, a thermal
eccentricity distribution, and a mass ratio distribution resulting
from (gravitationally-focused) random pairing of components from the
initial mass function.

\section{Method, initial conditions and $N$-body simulations}

To test our hypothesis that wide binary systems are formed during the
dissolution phase of star clusters, we carry out $N$-body simulations
using the STARLAB package \citep{spz2001}. For each star cluster were
draw $N$ stars from a \cite{kroupa2001} mass distribution in the range
$0.1-50\,$\msun. We study the properties of the resulting binary
population as a function of the number of member stars $N$ ($10\leq N \leq
1000$), the size $R$ of the star clusters (0.1~pc $\leq R \leq$ 1~pc),
and the primordial binary fraction $B$ ($0\%\leq B \leq 100\%$). We additionally
vary the virial ratio $Q\equiv E_K/E_p$ of the cluster, where $E_k$
and $E_p$ are the kinetic and potential energy of the cluster, respectively. We study the cases $Q=1/2$
(cluster in virial equilibrium) and $Q=3/2$ (expanding star
cluster). We carry out simulations for two different stellar density
distributions of the stars: (i) spherical \cite{plummer1911} models, and (ii)
models with substructured initial conditions with fractal parameter
$\alpha=1.5$ \citep[see][for details]{goodwin2004}. 

Simulations are carried out until the modeled cluster has completely
dissolved. At the end of each simulation we determine the properties
of the resulting binary population. We identify a pair of stars as a
binary system when (i) their binding energy is negative, and (ii) both
stars are each others mutual nearest neighbor. In our analysis we only
consider binary systems with a semi-major axis $a \leq 0.1$~pc, as
wider pairs are rather easily destroyed due to stellar encounters. We
also identify hierarchical multiple stellar systems ($\geq 3$ stars)
and impose the \cite{valtonen2008} stability criterion $a_{\rm
  out}/a_{\rm in} > Q_{st}$ for multiple stellar systems, where
$a_{\rm in}$ and $a_{\rm out}$ are the inner and outer orbits of a
(sub)system, respectively, and $Q_{st} \approx 3-10$ is a stability
parameter which depends on the orbital configuration. For wide
higher-order system with (outer) orbital periods of order $\sim
1$~Myr, this corresponds to a stability timescale of several billion
years.

\begin{figure}[tb]
  \centering
  \begin{tabular}{c}
    \includegraphics[width=1\textwidth,height=!]{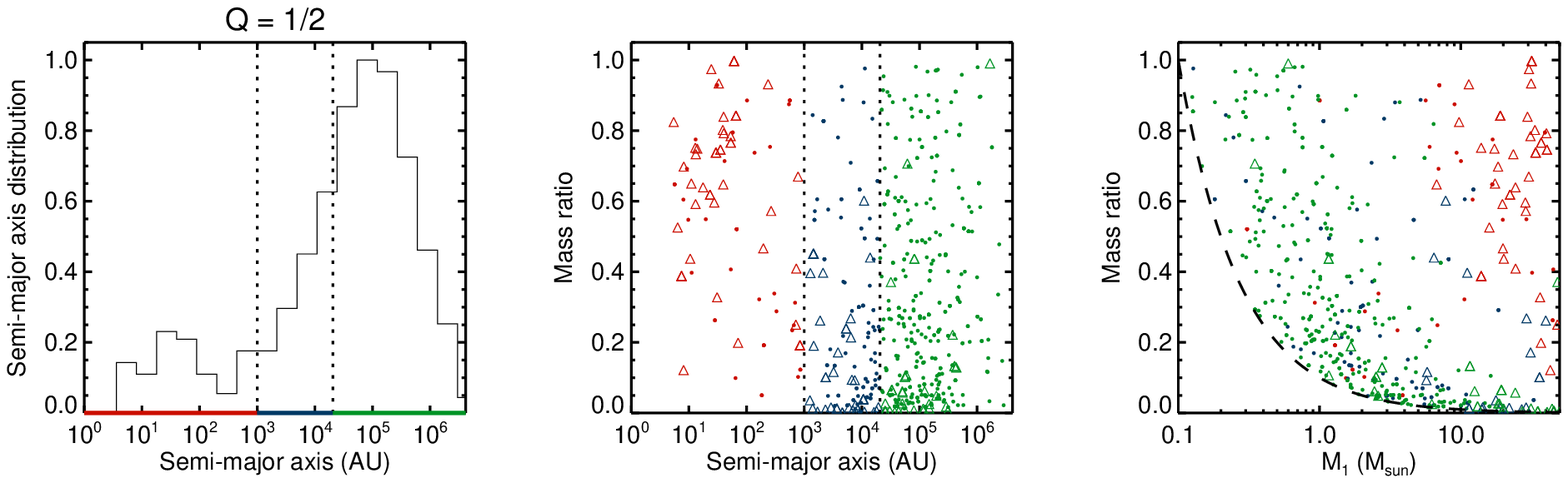}\\
    \includegraphics[width=1\textwidth,height=!]{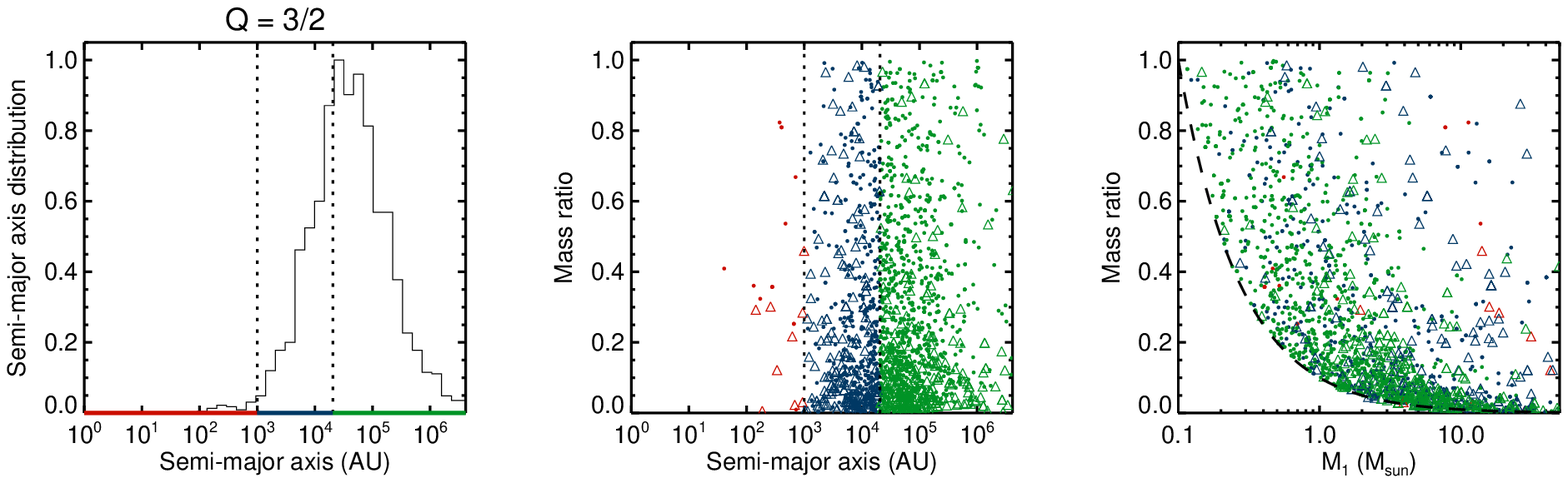}\\
  \end{tabular}
  \caption{The semi-major axis distribution ({\em left}), the
    correlation between mass ratio $q$ and semi-major axis $a$
    ({\em middle}) and between primary mass $M_1$ and mass ratio $q$
    ({\em right}). 
    The properties of the orbits of binary
    systems and higher-order multiple systems are indicated with the dots
    and triangles, respectively. For each multiple system with $n$
    stellar components, we
    have included all $n-1$ orbits. Results are shown for 50
    Plummer models with $N=1000$ and $R=0.1$~pc, and virial 
    ratios of $Q=1/2$
    ({\em top}) and $Q=3/2$ ({\em bottom}). The vertical dashed
    lines indicate $a=10^3$~AU and $a=0.1$~pc,
    respectively. The dashed curve in the right-hand panel
    indicates the minimum mass ratio $q_{\rm min}(M_1)= M_{\rm
      min} / M_1$.  \label{fig:kouwenhoven2}}
\end{figure}

\begin{figure}[tb]
  \centering
  \begin{tabular}{c}
    \includegraphics[width=1\textwidth,height=!]{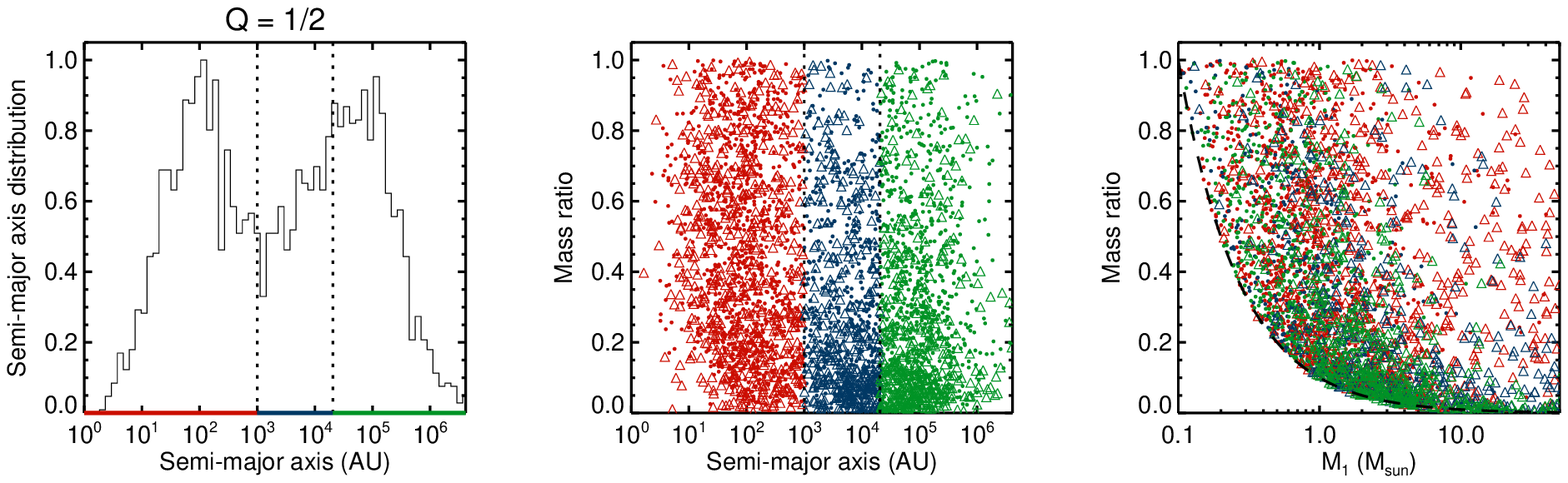}\\
    \includegraphics[width=1\textwidth,height=!]{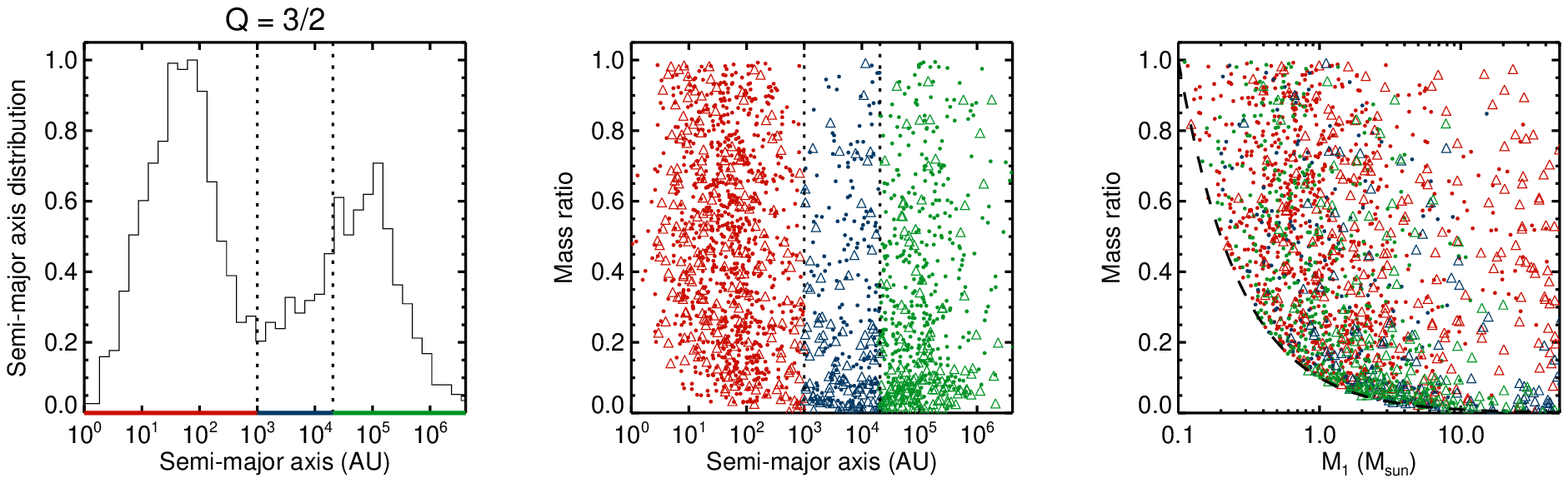}\\
  \end{tabular}
  \caption{Same as Fig.~\ref{fig:kouwenhoven2}, but now for substructured
    models (fractal parameter $\alpha=1.5$) with $N=1000$ and $R=0.1$~pc , and virial ratios 
    of $Q=1/2$ ({\em top}) and $Q=3/2$ ({\em bottom}); in each
    case fifty realisations have been simulated. \label{fig:kouwenhoven3}}
\end{figure}

\begin{figure}[tb]
  \begin{center}
    \includegraphics[width=0.7\textwidth, height=!]{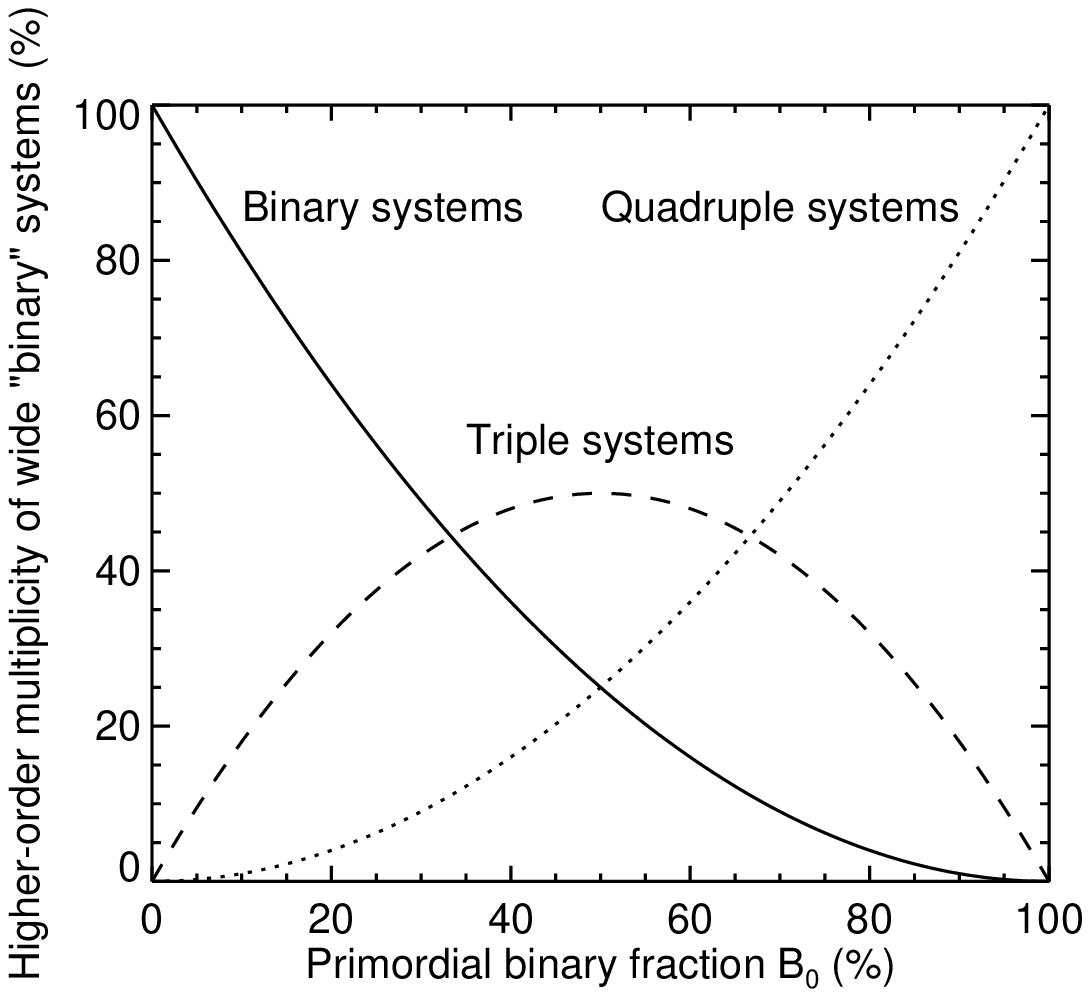}
    \caption{Most stars form as a member of a primordial binary system. Many wide
      ``binary'' systems formed during the star cluster dissolution process are
      therefore expected to be higher-order multiple systems. The
      relative multiplicities of wide systems can in principle be used to
      constrain the binary fraction $B_0$ at the time of star cluster
      dissolution.  \label{fig:kouwenhoven4}}
  \end{center}
\end{figure}

\section{The wide binary fraction, orbital elements and higher-order multiplicity} \label{section:elements}

A summary of the results of our $N$-body simulations is listed below. For an
extensive description of the results we refer to
\cite{kouwenhoven2010}. To illustrate the results, we show the
orbital properties of a selected sample of the simulations in
Figs.~\ref{fig:kouwenhoven2} and~\ref{fig:kouwenhoven3}. \\
\\
\noindent{\bf\em (a) The wide binary fraction.} After cluster
dissolution, the wide binary fraction (i.e., the fraction of wide
binary systems as compared to the total number of systems) ranges
between 1\% and 30\% for an individual star cluster. The exact value
depends the properties of the star cluster at the time of dissolution.
The structure of the star cluster at the moment of dissolution, as
well as the number of stars, affects the final number of wide binary
stars. Substructured star clusters (e.g., Fig.~\ref{fig:kouwenhoven3})
generate significantly more wide binary systems than spherical star
clusters (e.g., Fig.~\ref{fig:kouwenhoven2}). The wide binary fraction
increases with decreasing cluster membership and with increasing
initial virial ratio. The wide binary population in the Galactic field
($\sim 15\%$) results from a mixture of wide binary systems formed
from different types of star clusters. Its properties can therefore,
in principle, be
used to constrain the properties of young star clusters.\\
\\
\noindent{\bf\em (b) The semi-major axis distribution.} The resulting
semi-major axis distribution for wide binary systems is mainly in the
range $(0.1-1) R$, where $R$ is size of the star cluster at the moment
of dissolution. The semi-major axis distribution of the newly formed
binaries typically shows two peaks: a {\em dynamical peak} at small
values of $a$, resulting from three-body interactions, and an {\em
  dissolution peak} of wide binary systems formed during the
dissolution phase of the star cluster (this is clearly shown in the
left-hand panels in Fig.~\ref{fig:kouwenhoven3}). The ratio of the
number of binary stars in the {\em dynamical peak} and the {\em
  dissolution peak} depends on
the initial conditions of the star cluster (see above).\\
\\
\noindent{\bf\em (c) The eccentricity distribution.} The capture process
which results in wide binary formation is chaotic. The eccentricity
distribution for wide binary systems is therefore expected to be
thermal: $f(e) = 2e$
for $0\leq e < 1$ \citep{heggie1975}, which is confirmed by the simulations.\\
\\
\noindent{\bf\em (d) The mass ratio distribution.} The mass ratio
distribution for wide binary systems results from
gravitationally-focused random pairing \citep[e.g.,][and references
therein]{kouwenhoven2009} of the individual components. This implies
that wide binaries with a high-mass primary star have a small mass
ratio, while those with a low-mass binary have a high mass ratio (see
Figs.~\ref{fig:kouwenhoven2} and~\ref{fig:kouwenhoven3}). In addition,
the wide binary
fraction slowly increases with increasing primary star mass.\\
\\
\noindent{\bf\em (e) Orientation of the orbits.} The orientation of
the stellar spins of the two stars in a wide binary system are
randomly aligned. In the case of a wide multiple system, the orbital
orientations of the inner orbits are also randomly aligned with
respect to each other, and with respect to the orbit of the
wide orbit. \\
\\
\noindent{\bf\em (f) Implications for higher-order multiplicity.} A
significant fraction of star form as part of a primordial binary system. Both
components of a wide ``binary'' system are therefore expected to be
binary themselves. Recent observations suggest indeed that wide ``binary'' systems are
frequently triple or quadruple systems
\citep{makarov2008, mamajek2010, faherty2010, law2010}. The
multiplicity ratios among wide systems can therefore be used to
constrain the primordial binary fraction, or more specifically, the
binary fraction at the moment of star cluster
dissolution (see Fig.~\ref{fig:kouwenhoven4}).\\
\\
\noindent There is an ongoing debate about the wide binary formation
mechanism itself. \cite{kouwenhoven2010} shows that a pair of
(previously unbound) stars can form a wide binary system during the
dissolution phase, while the study of \cite{moeckel2011} shows that
a small (but transient) population of wide binary systems is
always present in a star cluster, and that this wide binary population is
frozen in when a star cluster dissolves. It may well be possible that
both mechanisms contribute to the formation of the wide binary
population in the Galactic field and halo.

\section{Summary}

Approximately 15\% of the known binary systems in the Galaxy have an
orbital separation larger than $10^3$~AU. These systems cannot be
primordial, simply because their orbital separations are comparable to
the size of young embedded clusters. Moreover, if they were able to
form in such environments, they would immediately be destroyed by
dynamical interactions with other stars. Dynamical capture in the
Galactic field or halo is highly improbable due to the low stellar
density and high velocity dispersion, and cannot explain the observed
wide binary population either.

We propose that wide binary systems form during the dissolution phase
of star clusters \citep[see][for details]{kouwenhoven2010}. In this
scenario, an escaping pair of stars with a small relative velocity can form a
wide binary system. $N$-body simulations confirm this hypothesis, and
allow us to predict the prevalence and orbital properties of the wide
binary population (\S~\ref{section:elements}), and the fraction of
triple and quadruple stars among
wide systems  (see Fig.~\ref{fig:kouwenhoven4}). These predictions can
be tested observationally, in particular those for the mass ratio
distribution and the higher-order multiplicity.

\acknowledgements 
M.B.N.K. was supported by the Peter and Patricia
Gruber Foundation through the IAU-PPGF fellowship, by the Peking
University One Hundred Talent Fund (985), and by the National Natural
Science Foundation of China (grants 11010237 and 11043007). The
authors acknowledge the Sheffield-Bonn Royal Society International
Joint Project grant, which provided financial support and the
collaborative opportunities for this work.

\bibliography{kouwenhoven}

\end{document}